\documentclass[twocolumn,showpacs,preprintnumbers,amsmath,amssymb,prl,superscriptaddress]{revtex4}
\usepackage{epsfig}
\newcommand{\bm}[1]{ \mbox{\boldmath $#1$}  }
\newcommand{\abs}[1]{\lvert #1 \rvert}
\newcommand{\pitwo}{ \frac{\pi}{2} }
\newcommand{\pisix}{ \frac{\pi}{6} }
\newcommand{\ud}{\,\mathrm{d}}
\newcommand{\rot}{{\cal R}}

\begin{document}

\title{Conditions for Efimov Physics for Finite Range Potentials}

\author{M. Th{\o}gersen}
\author{D.V. Fedorov}
\author{A.S. Jensen}
\affiliation{Department of Physics and Astronomy, University of Aarhus, 
DK-8000 Aarhus C, Denmark}

\author{B.D. Esry}
\author{Yujun Wang}
\affiliation{Department of Physics, Kansas State University, Manhattan, Kansas 66506, USA}

\date{\today}

\begin{abstract}
  We consider a system of three identical bosons near a Feshbach
  resonance in the universal regime with large scattering length
  usually described by model independent zero-range potentials.
  We employ the adiabatic hyperspherical approximation and derive the
  rigorous large-distance equation for the adiabatic potential for
  finite-range interactions.
  The effective range correction to the zero-range approximation must
  be supplemented by a new term of the same order. The non-adiabatic
  term can be decisive.
  Efimov physics is always confined to the range between effective
  range and scattering length. The analytical results agree with
  numerical calculations for realistic potentials.
\end{abstract}


\maketitle

\paragraph*{Introduction.}

Universal scaling properties in three-body systems arise when the
scattering length $a$ is much larger than the range $r_0$ of the
underlying two-body potential \cite{efimov73}. In this regime
certain three-body observables are universal in the sense that they
are model independent. This is colloquially referred to as
Efimov physics \cite{dincao05,kraemer06,braaten06,platter09}. Examples can be found in
nuclear systems, small molecules, and particularly in cold atoms where
the scattering length can be tuned to desired values using the
Feshbach resonance technique.

The universal scaling of Efimov trimers is usually said to exist for
rms-sizes between $r_0$ and $a$
\cite{efimov73,efimov91,braaten06,kraemer06}. The effective
range $R_e$ from a low-energy phase shift expansion is sometimes used
instead of $r_0$ in this statement
\cite{fedorov01,thogersen08-3,platter09}.  This ambiguity occurs
because $r_0$ and $R_e$ are often of the same order. However, for
narrow Feshbach resonances in atomic gases $R_e$ can be much larger
than $r_0$ \cite{bruun05}, and the implications for such systems
need to be explored.

Zero-range models, in particular in combination with the
hyperspherical approximation \cite{fedorov01,jonsell04,braaten06},
have been successful in semi-quantitative descriptions of three-body
systems in the universal regime. Semi-rigorous finite-range corrections have been
attempted by including the higher order
terms in the effective range expansion \cite{fedorov01,platter09} as a
step towards the full finite-range calculations as in
\cite{suno02,thogersen08-3} while maintaining the conceptual and technical
simplicity of the zero-range approximation.

The obvious generalization of the zero-range model is to substitute $-1/a$ with $-1/a+(R_e/2)k^2$,
where $k$ is the two-body wave-number, in the relevant expressions
for the logarithmic derivative of the total wave-function at small
separation of the particles. However, in three-body systems neither
the two-body wave-number nor the small separation are uniquely defined,
and rigorous inclusion of all terms of the given order is non-trivial.
The lack of rigor in previous works could have serious implications
for applications where finite-range effects are
important, such as the stability conditions for condensates in traps,
properties of cold atoms in lattices, and generally for Efimov
physics. Experimental progress \cite{kraemer06} will soon require this increased
accuracy near the boundaries of the universal regime.

In this Letter we derive, within the adiabatic hyperspherical
approximation \cite{nielsen01}, the rigorous asymptotic equation for
the adiabatic potential, which includes the finite-range correction
terms. The equation is suitable for the analytic studies of the
finite-range corrections in the three-boson problem. We investigate
the finite range corrections to the adiabatic potential and the non-adiabatic term
and compare with the zero-range approximation.

\paragraph*{Adiabatic eigenvalue equation.}

We consider three identical bosons of mass $m$ and coordinates $\bm
r_i$ interacting via a finite-range two-body potential $V$, where we
assume $V(r_{jk})=0$ for $r_{jk} = \abs{\bm r_j - \bm r_k} >r_0$.  Only
relative $s$-waves are included.  
We use the hyperradius $\rho^2=(r_{12}^2+r_{13}^2+r_{23}^2)2\mu/3$ and
hyperangles $\tan\alpha_i=(r_{jk}/r_{i,jk})\sqrt{3}/2$, where
$r_{i,jk}=\abs{\bm r_i-(\bm r_j+\bm r_k)/2}$ and $\mu$ is an arbitrary
parameter \cite{nielsen01}. In the following we shall use one set of
coordinates and omit the index.

The adiabatic hyperspherical approximation treats the hyperradius
$\rho$ as a slow adiabatic variable and the hyperangle $\alpha$ as the
fast variable. The eigenvalue $\lambda(\rho)\equiv\nu^{2}(\rho)-4$ of
the fast hyperangular motion for a fixed $\rho$ serves as the
adiabatic potential for the slow hyperradial motion. The eigenvalue is
found by solving the Faddeev equation for fixed
$\rho\ge\rho_c\equiv2r_0\sqrt{\mu}$,
\begin{equation} 
  \label{eq:faddeev}
  \Big[-\frac{\partial^2}{\partial\alpha^2}-\nu^2 +U\Big]\psi=-2U\rot[\psi].
\end{equation}
Here $\psi(\rho,\alpha)$ is the Faddeev hyperangular component,
\begin{equation}
  \label{eq:U}
  U(\rho,\alpha)=V(\rho\sin\alpha/\sqrt{\mu})\;m\rho^2/(\hbar^2\mu)
\end{equation}
is the rescaled potential, and
\begin{equation}
  \label{eq:rotation}
  \rot[\psi](\rho,\alpha)
  \equiv\frac{2}{\sqrt{3}}\int_{\abs{\frac{\pi}{3}-\alpha}}^{\frac{\pi}{2}-\abs{\frac{\pi}{6}-\alpha}}\psi(\rho,\alpha^{\prime})\ud\alpha^{\prime}
\end{equation}
is the operator that rotates a Faddeev component into another Jacobi
system and projects it onto $s$-waves. The total wave-function of the
three-body system is
$\Psi(\rho,\alpha)=f(\rho)\rho^{-5/2}\,\Phi(\rho,\alpha)$ where
\begin{equation}
  \label{eq:Phi}
  \Phi(\rho,\alpha)=\frac{\psi(\rho,\alpha)+2\rot[\psi](\rho,\alpha)}{\sin2\alpha}.
\end{equation}
The hyperradial function $f(\rho)$ satisfies the ordinary hyperradial
equation \cite{nielsen01} with the effective potential
\begin{equation}
  \label{eq:V_eff}
  V_\textup{eff}(\rho)=\frac{\hbar^{2}\mu}{m}\Big(\frac{\nu^{2}-\frac{1}{4}}{\rho^{2}}-Q\Big), \quad\,
  Q=\left\langle\Phi\right|\frac{\partial^{2}}{\partial\rho^{2}}\left|\Phi\right\rangle,
\end{equation}
where $Q$ is the non-adiabatic term and $\Phi$ is normalized to unity
for fixed $\rho$. 

We first divide the $\alpha$-interval $[0;\pi/2]$ into two regions: (I)
where $U \neq 0$, and (II) where $U = 0$.  The regions are separated at
$\alpha = \alpha_0$ where $\sin\alpha_0\equiv\sqrt{\mu}r_0/\rho =
\rho_c/(2\rho)$. In region (II) we have the free solution to Eq.~\eqref{eq:faddeev},
\begin{equation}
  \label{eq:phiII}
  \psi^{II}(\alpha)=N(\rho)\sin(\nu\alpha-\nu\pitwo)
\end{equation}
with the boundary condition,
$\psi^{II}(\pitwo)=0$ and normalization $N(\rho)$. In region (I), since $\alpha_0<\pi/6$,
Eq.~\eqref{eq:faddeev} simplifies to
\begin{equation}
  \label{eq:faddeev-like_eq}
  \Big[-\frac{\partial^2}{\partial \alpha^2} - \nu^2 + U \Big]\psi^{I}
  = -2 U \rot[\psi^{II}] ,
\end{equation}
with the solution $\psi^{I} = \psi^{Ih} - 2 \rot[\psi^{II}]$,
where $\psi^{Ih}$ and $-2\rot[\psi^{II}]$ are
homogeneous and inhomogeneous solutions, respectively. $\psi^{Ih}$ is the regular
solution to
\begin{equation}
  \label{eq:modif_angular_eq}
  \Big[
    -\frac{\hbar^2}{m}\frac{\partial^2}{\partial r^2}
    -\frac{\hbar^2 k_\rho^2}{m} 
    +V_\rho(r)
  \Big]\psi^{Ih}=0,
\end{equation}
\vspace{-1em}
\begin{equation} 
  \label{eq:potrho}
  V_\rho(r) \equiv V(\frac{\rho}{\sqrt{\mu}}\sin(\frac{\sqrt{\mu}}{\rho}r)),
\end{equation}
where $k_\rho=\sqrt{\mu}\nu/\rho$ and $r=\alpha\rho/\sqrt{\mu}$. When $\alpha\to\alpha_0$,
\begin{equation}
  \label{eq:phiI}
  \psi^{Ih}\propto \sin(k_\rho r+\delta_\rho),
\end{equation}
where the modified phase shift $\delta_\rho(k_\rho)$
arises from the modified two-body potential, $V_\rho$.
The solutions $\Phi$ in region (I) and (II) are now matched smoothly, leading to
\begin{equation}
  \label{eq:logderiv_match}
  \frac{\partial}{\partial\alpha}\ln\psi^{Ih}\big|_{\alpha_0}=
  \frac{\partial}{\partial\alpha}\ln\big(\psi^{II}+2\rot[\psi^{II}]\big)\big|_{\alpha_0}.
\end{equation}
After inserting Eqs.~\eqref{eq:phiII} and \eqref{eq:phiI}, this
equation becomes
\begin{equation}
  \label{eq:generalized_efimov_eq}
  \frac{\sqrt{\mu}}{\rho} \ \frac{ -\nu\cos(\nu\pitwo)+ \frac{8}{\sqrt{3}}\sin(\nu\pisix) }
  {\sin(\nu\pitwo)} 
  = k_\rho\cot\delta_\rho(k_\rho),
\end{equation}
which defines $\nu$ as function of $\rho$. 
The right-hand-side deviates from the zero-range approximations
\cite{jonsell04,platter09} by using the rigorously defined phase
shifts $\delta_\rho$ for $V_\rho$ instead of the original phase shifts
$\delta$.

\paragraph*{Effective range expansion.}

\begin{figure}[tb]
  \epsfig{file=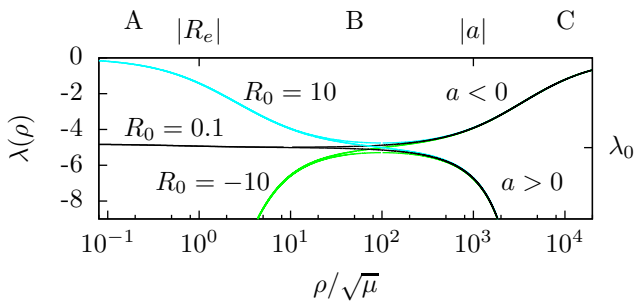, scale=1}
  \caption{(color online) Adiabatic eigenvalues $\lambda(\rho)$ from
    Eq.~\eqref{eq:generalized_efimov_eq2} as function of hyperradius
    $\rho$, for large scattering length $a$ and negative $R_e$.
    Different values of the model dependent length $R_0$ are used,
    showing that the universal solution $\lambda_0$ exists in the
    region $\abs{R_0}\ll\rho/\sqrt{\mu}\ll\abs{a}$. Lengths are in
    units of $\abs{R_e}$.}
  \label{fig:lambda-analytic}
\end{figure}

In the limit $\rho\gg \rho_c$, the $\rho$-dependent potential, $V_\rho(r)$,
approaches $V(r)$, and consequently
$\delta_\rho$ approaches $\delta$.  
The $\rho$-dependent low-energy effective range expansion corresponding to $V_\rho$ is
then to second order
\begin{equation} 
  \label{eq:effexp}
  k_\rho\cot\delta_\rho(k_\rho)\Big|_{k_\rho\to0}
  \approx
-\frac{1}{a(\rho)}+\frac{R_e(\rho)}{2}k_\rho^2,
\end{equation}
where $a(\rho)$ and $R_e(\rho)$ are functions of $1/\rho^2$ that
converge to $a$ and $R_e$ for $\rho\to\infty$.
Up to $1/\rho^2$ in Eq.~\eqref{eq:effexp} we get
\begin{equation}  \label{eq:scatexp}
      \frac{1}{a(\rho)} \approx\frac{1}{a} +
      R_V \frac{\mu}{\rho^2},\quad
      R_e(\rho)\approx R_e.
\end{equation}
The model dependent expansion parameter $R_V$, or ``scattering length
correction'', is found to be
\begin{equation}
  \label{eq:c02_integral_rep2}
  R_V 
  =\frac{m}{6\hbar^2}\langle V'r^3 \rangle_u
  =\frac{m}{6\hbar^2}\int_0^{r_0}V'(r)r^3 u(r)^2\ud r,
\end{equation}
where $u$ is the zero-energy two-body radial wavefunction,
asymptotically equal to $1-r/a$. Eq.~\eqref{eq:generalized_efimov_eq}
then becomes
\begin{equation}
  \label{eq:generalized_efimov_eq2}
  \frac{\sqrt{\mu}}{\rho} \ \frac{ -\nu\cos(\nu\pitwo)+ \frac{8}{\sqrt{3}}\sin(\nu\pisix) }
  {\sin(\nu\pitwo)} 
  = -\frac{1}{a}+\frac{R_e}{2}\frac{\nu^2\mu}{\rho^2}-\frac{R_V \mu}{\rho^2}.
\end{equation}
This equation without the last two finite range terms has the
well-known purely imaginary solution $\nu_0=1.00624 i$, or
$\lambda_0=-5.0125$, for $\rho/\sqrt{\mu}\ll\abs{a}$. This solution
gives $V_\textup{eff}\propto-1/\rho^2$ which is the basis of Efimov
physics. The $R_e$-term was included in \cite{fedorov01,platter09},
but not the model dependent $R_V$-term. The latter term makes the finite-range
corrections to the zero-range adiabatic eigenvalues explicitly
non-universal. The last two terms in
Eq.~\eqref{eq:generalized_efimov_eq2} restrict the solution
$\lambda_0$ to the region $\abs{R_0} \ll \rho/\sqrt{\mu} \ll \abs{a}$,
where
\begin{equation}
  \label{eq:RE}
  R_0 =\frac{R_e}{2}\nu_0^2-R_V,
\end{equation}
as seen in Fig.~\ref{fig:lambda-analytic} where the lowest solution to
Eq.~\eqref{eq:generalized_efimov_eq2} is shown for different parameter
choices. Thus, naively one would think that the lower limit for
Efimov physics is determined by the model dependent length
$\abs{R_0}$. However, we will show later that $Q$ restores
universality and recovers the model independent effective range,
$R_e$.

\begin{figure}[tb]
  \epsfig{file=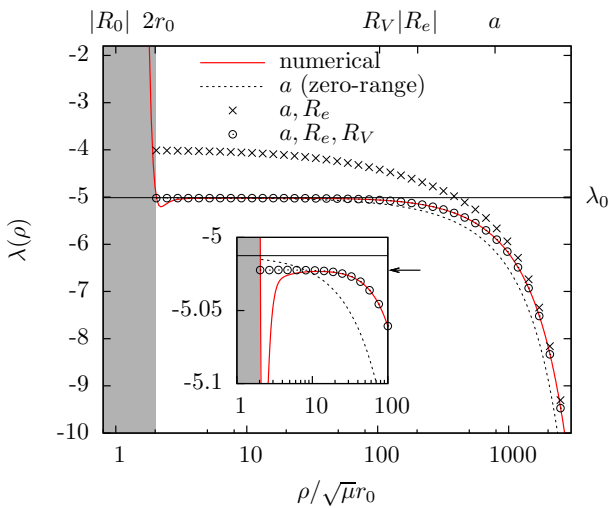,scale=0.9}
  \caption{(color online) Exact numerical adiabatic eigenvalues
    $\lambda(\rho)$ for a potential with barrier,
    Eq.~\eqref{eq:potential-with-barrier} (solid red line), compared
    to solutions of the eigenvalue equation,
    Eq.~\eqref{eq:generalized_efimov_eq2}.  The zero-range model
    (dotted line) includes only $a$.  Crosses include $a,R_e$-terms
    and circles include $a,R_e,R_V $-terms. The inset shows details
    around $\lambda_0$. The arrow indicates the effect of the
    correction $\lambda_0-2R_0 /R_e$.}
  \label{fig:lambda-numerics}
\end{figure}

First, to illustrate the necessity of both $1/\rho^2$-terms in
Eq.~\eqref{eq:generalized_efimov_eq2} we consider a large negative
effective range corresponding to a narrow Feshbach resonance
\cite{bruun05}. To model the large $\abs{R_e}$ we pick an attractive
potential with barrier
\begin{equation}
  \label{eq:potential-with-barrier}
    V(r)=D\ \text{sech}^2\Big(\chi\frac{r}{r_0}\Big)
        +B \exp\Big(-2(\chi\frac{r}{r_0}-2)^2\Big),
\end{equation}
where $D=-138.27$, $B=128.49$ in units of $\hbar^2/(mr_0^2)$, and
$\chi=4.6667$. The potential is negligible outside the range $r_0$.
The low-energy parameters are $a=556.88 $, $R_e=-142.86 $, $R_V
=73.031 $, and $R_0 =-0.71 $ in units of $r_0$.  In
Fig.~\ref{fig:lambda-numerics} we compare $\lambda(\rho)$ obtained by
exact numerical solution of the Schr{\"o}dinger equation \cite{suno02}
containing the interaction Eq.~\eqref{eq:potential-with-barrier} with
the solution of Eq.~\eqref{eq:generalized_efimov_eq2}.
In the zero-range model (including only $1/a$), the $-\rho^2$
divergence for large $\rho$ is below the numerical solution. At small
distances, $\lambda$ approaches $\lambda_0$, above the numerical
solution.  Inclusion of the $R_e$-term, as in
\cite{fedorov01,platter09}, provides a better large-distance behavior
(since the dimer binding energy is corrected), but overshoots
dramatically for $\rho/\sqrt{\mu} \lesssim a$ by approaching
$\lambda=-4$.  Including consistently both $R_e$- and $R_V$-terms
leads to complete numerical agreement with the exact numerical
solution except for very small $\rho$-values where higher order terms
are needed in Eq.~\eqref{eq:generalized_efimov_eq2}.

\paragraph*{Non-adiabatic corrections.}

We shall show that the non-adiabatic term restores model independence
and recovers $\abs{R_e}$ as the limit for the region of Efimov
physics. For simplicity we only consider the limit $\abs{a}=\infty$ 
and assume $\abs{R_0}\ll\abs{R_e}$.  We first
consider $\rho/\sqrt{\mu}\ll\abs{R_e}$ (region A in
Fig.~\ref{fig:lambda-analytic}).  Expansion of
Eq.~\eqref{eq:generalized_efimov_eq2} to first order in $(\nu-\nu_0)$
gives a small constant correction
\begin{equation}
  \label{eq:nu_efimov_large_Re}
  \nu=\nu_0-\frac{R_0}{\nu_0 R_e}\Big(1+O(\frac{\rho}{R_e})\Big).
\end{equation}
This correction is marked by the arrow in
Fig.~\ref{fig:lambda-numerics} (it is out of the range of
Fig.~\ref{fig:lambda-analytic}).  This gives
\begin{equation}
  \label{eq:V_eff_efi_large_Re}
  V_\textup{eff}(\rho)=\frac{\hbar^2\mu}{m} 
  \Big(\frac{\nu_0^2-1/4-2R_0 /R_e}{\rho^2} - Q\Big).
\end{equation}
To evaluate $Q$ we note that a large negative effective range (for
$\abs{a}=\infty$) implies that the two-body wavefunction $u$ is
localized mainly inside the potential range. Then the angular
three-body wavefunction $\Phi$ can be approximated by
$u/\sin(2\alpha)$. The result is $Q=c/\rho^2$, where $c\simeq -5/4$
as confirmed numerically.  This term cancels the main $1/\rho^2$-part
in Eq.~\eqref{eq:V_eff_efi_large_Re} and hence prohibits Efimov
physics for $\rho/\sqrt{\mu}\ll\abs{R_e}$.  The intuitive reason is
that the two-body wavefunction is essentially zero outside the
potential, despite the large scattering length, and hence three
particles can not interact at large distances.

When $\rho/\sqrt{\mu}\gg\abs{R_e}$ (region B in
Fig.~\ref{fig:lambda-analytic}) we find
\begin{equation}
  \label{eq:nu_efimov}
  \nu=\nu_0+\nu_0 c_0 \frac{R_0 \sqrt{\mu}}{\rho}\Big(1+O(\frac{R_e}{\rho})\Big),
\end{equation}
\vspace{-1em}
\begin{equation}
  \label{eq:kE}
    c_0=\frac{\sin(\nu_0\pitwo)/\nu_0}{\frac{4\pi}{3\sqrt{3}}\cos(\nu_0\pisix)
      -\cos(\nu_0\pitwo)+\nu_0\pitwo\sin(\nu_0\pitwo)},
\end{equation}
or $c_0\simeq -0.671$. This gives
\begin{equation}
  \label{eq:V_eff_efi}
    V_\textup{eff}(\rho)=\frac{\hbar^2\mu}{m} \Big(\frac{\nu_0^2-1/4}{\rho^2}
    +\frac{c_0\nu_0^2\sqrt{\mu}}{\rho^3}
    \left( R_e\nu_0^2-2R_V \right) - Q\Big).
\end{equation}
The $1/\rho^3$ dependence of the correction to the Efimov potential
$1/\rho^2$ was expected \cite{efimov91}. The model independent term
proportional to $R_e/\rho^3$ was recently
calculated in \cite{platter09}. However, we also get a model
dependent term $R_V/\rho^3$ which is of the same order.
$Q$ generally receives contributions both from distances inside and
outside the finite-range potential. Zero-range models only have the
external part of the wavefunction, which depends on $\rho$ only though
the eigenvalue $\nu(\rho)$. The zero-range result for $Q$ is then
\begin{equation}
  Q_\textup{ZR} =M_0(\frac{\partial\nu}{\partial\rho})^2 
  =M_0 c_0^2 \nu_0^2 \frac{R_0^2\mu}{\rho^4},
\end{equation}
where $M_0=\langle
\Phi|\partial^2\Phi/\partial\nu^2\rangle|_{\nu=\nu_0}$. This fourth
order correction can be neglected in Eq.~\eqref{eq:V_eff_efi}, as was
done in \cite{platter09}. However, the internal part of the
wavefunction contributes to order $1/\rho^3$.
To estimate this $1/\rho^3$-term we take the analytically solvable
finite square well potential of range $r_0$ and $\abs{a}=\infty$. This
fixes $R_e=r_0$ and $R_V=n^2\pi^2r_0/24$ where $n$ is the number of
bound states (including the zero-energy state).  We find
\begin{equation}
  \label{eq:Qbox}
  Q_\textup{box}=c_0\nu_0^2(\frac{R_e}{2}-2R_V)\frac{\sqrt\mu}{\rho^3},
\end{equation}
neglecting $1/\rho^4$-terms. The model dependent $R_V$-terms in
Eqs.~\eqref{eq:V_eff_efi} and \eqref{eq:Qbox} cancel exactly, giving
\begin{equation}
  \label{eq:V_eff_efi_box}
    V_\textup{eff}^\textup{box}(\rho)=\frac{\hbar^2\mu}{m} \Big(
    \frac{\nu_0^2-1/4}{\rho^2}
    +c_0\nu_0^2(\nu_0^2-\frac{1}{2})\frac{\sqrt\mu R_e}{\rho^3}
    \Big).
\end{equation}
So the effective potential receives a $R_e/\rho^3$ correction where
the model dependent coefficient is different from zero-range models
\cite{platter09} because of the inclusion of $Q$.  We also expect the
$R_V$-terms to cancel for general potentials. In conclusion, the
Efimov effect persists for $\rho/\sqrt{\mu}\gg\abs{R_e}$.

\paragraph*{Atom-dimer potential.}

We have seen that model dependent corrections to $\lambda_0$ are
cancelled by equivalent terms in $Q$. A similar effect occurs 
for the atom-dimer channel potential.
Suppose the binding energy
is $B_D=\hbar^2k_D^2/m$ with corresponding wave number $k_D>0$. Then
$\nu = i k_D \rho/\sqrt{\mu}$ is an asymptotic solution to
Eq.~\eqref{eq:generalized_efimov_eq} and $\lambda$ diverges as
$-\rho^2$ corresponding to a bound dimer and a free particle. For this
solution, the effective range expansion Eq.~\eqref{eq:effexp} does not
hold, since asymptotically $k_\rho\to ik_D$ is finite. Instead
Eq.~\eqref{eq:modif_angular_eq} reduces to the radial two-body
equation, with a normalized bound state $s$-wave function $u_D(r)$.
Treating $V_\rho-V\propto 1/\rho^2$ as a perturbation gives the
correction
\begin{equation}
  \label{eq:lambda_AD-corrections}
  \frac{\lambda+4}{\rho^2}=-\frac{k_D^2}{\mu}
  -\int_0^\infty \!\!\! r^3 u_D^2\frac{mV'(r)}{6\hbar^2} \ud r \frac{1}{\rho^2} 
  + O(\frac{1}{\rho^4}).
\end{equation}
Since $\rot[\psi]$ is exponentially small for the atom-dimer solution,
$Q$ can be computed using the unperturbed wavefunction
$\psi=\sqrt{\rho}u_D(\alpha\rho/\sqrt{\mu})$, giving
\begin{equation}
  \label{eq:Q00_AD}
  Q=-\frac{1}{4\rho^2}
  +\int_0^\infty \!\! u_D(ru_D'+r^2u_D'')\ud r \frac{1}{\rho^2}  + O(\frac{1}{\rho^4}).
\end{equation}
By using the two-body radial equation and partial
integration the two integrals in Eqs.~\eqref{eq:lambda_AD-corrections}
and~\eqref{eq:Q00_AD} cancel. Thus the $1/\rho^2$-terms in the
effective potential Eq.~\eqref{eq:V_eff} cancel exactly,
giving $V_\textup{eff}(\rho)=-B_D$ up to order $1/\rho^4$. Thus
$V_\textup{eff}$ only depends on $R_e$ through $B_D$.

\paragraph*{Effective range for Feshbach resonances.}
The effective range near a Feshbach resonance has been estimated using
a coupled-channels zero-range model \cite{bruun05}, as
\begin{equation}
  \label{eq:Re_feshbach}
  R_e=-2(  \Delta B\,  m\, \Delta\mu\,  a_{bg})^{-1},
\end{equation}
where $\Delta B$ is the magnetic field width, $\Delta\mu$ is the
magnetic moment difference between the channels, and $a_{bg}$ is the
background scattering length. As an example we take the alkali atoms
$^{39}$K with the very narrow Feshbach resonance at $B=825$G having
parameters $\Delta B=-32$mG, $\Delta\mu=-3.92\mu_B$, and
$a_{bg}=-36a_0$ \cite{derrico07}. This gives the large effective range
$R_e=-2.93\times 10^4 a_0$.  For $^{39}$K, $r_0$ is of the order of
the van der Waals length $l_\textup{vdW}=1.29\times 10^2a_0$
\cite{braaten06}.  Since $\abs{R_e}\gg r_0$, $\abs{R_e}$ determines
the lower limit for Efimov physics and corrections to the universal
regime are of order $R_e/a$ (not $l_\textup{vdW}/a$). Thus, the window
for universal physics is reduced.

\paragraph*{Summary and Conclusions.}

We consider a three-body system of identical bosons with large
scattering length modelling a Feshbach resonance. The Efimov physics
occurring in this ``universal regime'' is customarily accounted for by
zero-range models. We use the adiabatic hyperspherical approximation
and derive rigorously a transcendental equation to determine the
asymptotic adiabatic potential for a general finite-range potential.
We solve this equation for large scattering length, investigate finite
range effects, and compare with exact numerical results.

Inclusion of the effective range correction to the adiabatic
potential is insufficient in general.  Crucial corrections of the same
order must also be included from both the scattering length and the
non-adiabatic term.  These two contributions may separately be large
but they tend to cancel each other.
Accurate results in zero-range models must account for these new
corrections.
In conclusion, the window for Efimov physics is precisely open between
the effective range (not the potential range) and the scattering
length.

\paragraph*{Acknowledgements.}
This work was done partly in the framework of the Nordforsk Network on
Coherent Quantum Gases. BDE and YW acknowledge support from the
U.S. National Science Foundation and the U.S. Air Force Office of
Scientific Research.

\bibliographystyle{h-physrev}

\end{document}